\documentclass[a4paper,11pt]{article}
\pdfoutput=1 

\usepackage{jheppub} 

\usepackage[T1]{fontenc} 
\usepackage{amsmath}
\usepackage{amssymb}
\usepackage{amsfonts}
\usepackage{graphicx}
\usepackage{enumitem}
\usepackage{bm}
\usepackage{rotating}
\usepackage{hyperref}
\usepackage{pbox}
\usepackage[dvipsnames]{xcolor}
\usepackage{array}
\usepackage{physics}
\usepackage{dsfont}
\usepackage{mathtools}
\usepackage{leftidx}
\usepackage{lmodern}
\usepackage[normalem]{ulem}
\usepackage{braket}
\usepackage{tensor}
\usepackage{mathrsfs}
\usepackage{float}
\usepackage[title]{appendix}

\usepackage{tikz}
\usetikzlibrary{arrows.meta,decorations.pathmorphing,math}

\numberwithin{equation}{section}

\newcommand{\be}{\begin{equation}}
\newcommand{\ee}{\end{equation}}
\newcommand{\ba}{\begin{eqnarray}}
\newcommand{\ea}{\end{eqnarray}}

\newcommand{\beq}{\begin{equation}}
\newcommand{\eeq}{\end{equation}}
\newcommand{\beqa}{\begin{eqnarray}}
\newcommand{\eeqa}{\end{eqnarray}}
\newcommand{\nn}{\nonumber}

\title{Generalized Lense--Thirring metrics:  higher-curvature corrections and solutions with matter}

\date{January 30, 2021}

\author[a,b]{Finnian Gray,}
\author[b,c]{Robie A. Hennigar,}
\author[a,b]{David Kubiz\v n\'ak,}
\author[b,a]{\\Robert B. Mann,}
\author[a,b]{Manu Srivastava}

\affiliation[a]{Perimeter Institute for Theoretical Physics, Waterloo, Ontario N2L 2Y5, Canada}

\affiliation[b]{Department of Physics and Astronomy, University of Waterloo, Waterloo, Ontario, N2L 3G1, Canada}
\affiliation[c]{
	Department of Physics and Computer Science, Wilfrid Laurier University, 
	Waterloo, Ontario, Canada N2L 3C5
}

\emailAdd{fgray@perimeterinstitute.ca}
\emailAdd{rhennigar@uwaterloo.ca}
\emailAdd{dkubiznak@perimeterinstitute.ca}
\emailAdd{rbmann@uwaterloo.ca}
\emailAdd{msrivastava@perimeterinstitute.ca}

\abstract{The Lense--Thirring spacetime describes a 4-dimensional slowly rotating approximate solution of vacuum Einstein equations valid to a linear order in rotation parameter. It is fully characterized by a single metric function of the corresponding static (Schwarzschild) solution. In this paper, we introduce a generalization of the Lense--Thirring spacetimes to the higher-dimensional multiply-spinning case, with an ansatz that is not necessarily fully characterized by a single (static) metric function. This generalization lets us study slowly rotating spacetimes in various higher curvature gravities as well as in the presence of non-trivial matter. Moreover, the ansatz can be recast  in Painlev{\'e}--Gullstrand form (and thence is manifestly regular on the horizon) and admits a tower of exact rank-2 and higher rank Killing tensors that rapidly grows with the number of dimensions. In particular, we construct slowly multiply-spinning solutions in Lovelock gravity and notably show that in four dimensions Einstein gravity is the only non-trivial theory amongst all up to quartic curvature gravities that admits a Lense--Thirring solution characterized by a single metric function.       
}

\begin{document} 
\maketitle
\flushbottom

\section{Introduction}
The problem of understanding the gravitational field of a {\em rotating body} is one of great physical and theoretical significance. 
From an {\em astrophysical viewpoint}, a tremendous  amount of energy is stored in the spin of a body, which in turn can drive a variety of astrophysical  phenomena~\cite{Reynolds:2019uxi}.  For example, the spin of a central supermassive black hole can energize powerful jets for hundred of millions of years in active galactic nuclei before becoming exhausted. These jets in turn deposit their energy within their host galaxies,  disrupting gas and inhibiting   star formation~\cite{Fabian:2012xr}.  This has now become the standard  model for how active galactic nuclei  jets  are driven~\cite{Blandford:1977ds}, making black hole spin integral in galaxy formation.  Another example is that of the formation of black holes: whether supermassive black holes form from disk-accretion or  mergers of smaller black holes can in part be determined by the fraction of their population that is rapidly spinning~\cite{Volonteri:2004cf}. 
 
From a theoretical stance, any spinning body induces an effect on spacetime known as {\em frame-dragging} in which, for example  the orbit of a nearby spinning test particle precesses about the massive rotating object. It is  the  gravitational analogue of electromagnetic induction, and was first predicted by Lense and Thirring in 1918~\cite{lense1918influence}, and first measured in 2008~\cite{Everitt:2011hp,Everitt:2015qri}.  Another familiar, albeit extreme, example of frame dragging is the existence of an ergosphere around rotating black holes, a region where the dragging is so strong that every object has to co-rotate with the black hole. Such a region gives rise to superradiant phenomena, e.g.  \cite{Brito:2015oca}.
Rotation is likewise important in black hole thermodynamics~\cite{Kubiznak:2016qmn}, in which the angular momentum and angular velocity of the event horizon of a black hole play a significant role in determining the phase behaviour and thermodynamic stability of these objects~\cite{Kubiznak:2012wp,Altamirano:2013ane,Altamirano:2014tva}.

Although a number of exact solutions  with rotation are known in Einstein gravity~\cite{kerr1963gravitational,Newman:1965tw,Myers:1986un,Gibbons:2004js, Sen:1992ua, Chong:2005hr}, for general theories of gravitation (or in the presence of various matter fields) they are quite difficult to obtain. On the other hand slowly rotating solutions are somewhat easier to come by. Perhaps the nicest example is that of Lovelock gravity, where the slowly rotating solution for a (single) rotation parameter has been known for some time~\cite{Kim:2007iw, Camanho:2013pzg}, and the equations of motion can be solved exactly.\footnote{Attempts to analytically push beyond the slowly rotating regime have generally been met with failure, except in certain special cases~\cite{Anabalon:2009kq, Ett:2011fy, Cvetic:2016sow}, though numerical studies suggest the full rotating solutions exist~\cite{Brihaye:2010wx}.} Beyond Lovelock gravity, other interesting higher-curvature theories have been the focus of some study -- examples appear in Einstein Gauss--Bonnet gravity in $d=4$ and $d=5$~\cite{Konoplya:2020fbx}.
Moreover, the slowly rotating solutions of Einsteinian Cubic Gravity were studied in~\cite{Adair:2020vso}, while the case of five-dimensional cubic and quartic quasi-topological gravities were studied in~\cite{Fierro:2020wps}, allowing for two independent rotation parameters in five dimensions.  These cases are interesting examples where theories that generally have fourth-order equations of motion reduce to second-order equations of motion for a particular case of interest. However, in each case the slowly rotating solutions must be obtained numerically, and are more complicated than the corresponding Einstein gravity solutions. More general solutions have been obtained in the context of four-dimensional effective field theory, see e.g.~\cite{Pani:2011gy,Cardoso:2018ptl,Cano:2019ore,Buoninfante:2020qud}, and dynamical Chern--Simons gravity \cite{Alexander:2021ssr, Srivastava:2021imr}. 
Even within the realm of Einstein-Maxwell gravity,  exact charged solutions are not known in higher dimensions, or even in four dimensions when   non-linear generalizations of   Maxwell's theory are considered. 
Here one of our aims is to study slowly rotating solutions in more general theories of gravity, and also with matter.

The prototypical solution for Einstein gravity in four dimensions is the well-known
Lense--Thirring metric \cite{lense1918influence}, which reads
\begin{align} 
ds^2 &= -f dt^2 + \frac{dr^2}{f} + 2 a (f-1) \sin^2 \theta d t d \phi + r^2(\sin^2\theta d\phi^2+d\theta^2) \, ,\nonumber
\\
f &= 1 - \frac{2M}{r} \, .
\label{LTmet}
\end{align}
It solves the vacuum Einstein equations to leading order in the rotation parameter $a$,  and describes spacetime outside of a rotating body. The metric also arises as the slow rotation ($a \to 0$) limit of the Kerr metric \cite{kerr1963gravitational}.  
 
A simple observation about the Lense--Thirring metric \eqref{LTmet} is that it is completely characterized by the static Schwarzschild solution: the metric component $g_{t \phi}$ is written in terms of the static metric function $f$. In the context of four-dimensional Einstein gravity, this point may be understood in relation to the  Newman--Janis trick~\cite{Newman:1965tw}. The fact that the full Kerr solution can be generated from the `seed' static solution of course implies that the Lense--Thirring metric is obtained when the trick is truncated at $\mathcal{O}(a)$. 

What is perhaps a bit more interesting is that the same holds true in higher dimensions, including rotations in multiple planes. In this case,  the  general $d$-dimensional Lense--Thirring metric reads \cite{Myers:1986un}
\be\label{LTHDein}
ds^2=-fdt^2+\frac{dr^2}{f}+\sum_{i=1}^m \mu_i^2 a_i (f-1) dt d\phi_i +r^2(\sum_{i=1}^m d\mu_i^2+
\mu_i^2 d\phi_i^2) + \epsilon r^2 d\nu^2\,,
\ee
for  $m=\lfloor\frac{d-1}{2}\rfloor$ independent rotation parameters $a_i$, where $\lfloor A \rfloor$ denotes the whole part of $A$, and
\be 
f = 1 - \frac{16 \pi M}{(d-2) \Omega_{d-2} r^{d-3}}  
\ee
with  $M$   the mass of the rotating body, and $\Omega_{d}$ the volume of the $d$-dimensional sphere. In the above, the coordinates $\mu_i$ and $\nu$ obey the following constraint:
\be\label{constr}
\sum_{\mu=1}^m \mu_i^2+\epsilon \nu^2=1\,,
\ee
where $\epsilon=1,0$ in even, odd dimensions. The metric \eqref{LTHDein} solves the $d$-dimensional Einstein equations to $\mathcal{O}(a)$ in the rotation parameter, and is still characterized completely by the static metric function. The same remains true even in the presence of a cosmological constant $\Lambda$, where $f$ above is replaced with\footnote{ This higher-dimensional case is less obviously equivalent to the Newman--Janis trick, as, to the best of our knowledge, it is not yet known whether such a trick exists for arbitrary rotations in arbitrary dimensions -- see~\cite{Xu:1988ju} for the singly spinning Kerr metric in all dimensions, and~\cite{Erbin:2014lwa} for the general five dimensional Myers--Perry case, and a discussion of the problems of further generalization. Nonetheless, it is clearly in the same spirit.} \cite{Gibbons:2004js}: 
\be\label{fLambda}
 f = 1 - \frac{16 \pi M}{(d-2) \Omega_{d-2} r^{d-3}}+\frac{r^2}{\ell^2}\,,\quad \Lambda=-\frac{(d-1)(d-2)}{2\ell^2}\,,
\ee
and $\ell$ is known as the AdS radius\footnote{This works equally for positive cosmological constant, sending  $\ell\to i\ell$, i.e. for asymptotically de Sitter black holes.}.

It is natural then to wonder how general this feature is. That is, how generic (or special) is it for the Lense--Thirring metric to be completely determined by the static metric?  Here we  address this question in the context of  i) vacuum solutions of higher curvature gravity and ii) solutions with matter in Einstein gravity. To do this, we shall consider the following generalization of the Lense--Thirring spacetime  (and its variant discussed in the next section):
\be\label{LTHD}
ds^2=-N fdt^2+\frac{dr^2}{f}+\sum_{i,j=1}^m  \mu_i^2p_{ij} a_j dt d\phi_i  +r^2(\sum_{i=1}^m d\mu_i^2 +\mu_i^2 d\phi_i^2)+r^2\epsilon d\nu^2 \,, 
\ee
where 
$f, N, p_{ij}$ are  functions of the radial coordinate $r$ (we assume that the lapse satisfies $N > 0$, at least in the exterior), and coordinates $\mu_i$ are $\nu$ are as before. 

We shall demonstrate that  the following  ``Einstein-like'' form:
\be\label{pi_Einstein}
N=1\,,\quad p_{ij}=(f-1)\delta_{ij}\,,
\ee 
is very special, and will not generically survive 
when higher-curvature terms and/or matter are added to the action. In fact, we shall prove the following theorem:\\
\begin{quote}
{\bf Theorem.}\emph{ In $d=4$ dimensions,  Einstein gravity is the only nontrivial gravitational theory up to  powers quartic in the Riemann curvature tensor, whose (vacuum) generalized Lense--Thirring solutions \eqref{LTHD} are of the form \eqref{pi_Einstein}.}
\end{quote}

On the other hand, as we shall see, in higher dimensions there are other nontrivial  theories that admit solutions of the form \eqref{pi_Einstein}. Examples include Lovelock gravity in all dimensions, and a subset of quartic generalized quasi topological gravities~\cite{Hennigar:2017ego} in five dimensions. When   matter is included, one easily finds examples with more general $N$ or $p_{ij}$. As we shall see in Sec.~\ref{Sec:matter}, an especially interesting example is that of a slowly rotating black hole in minimal gauged supergravity where the ``Chern--Simons'' term `mixes' rotation parameters in various rotation 2-planes, resulting in distinct $p_{ij}$ in each of the planes.

Before we discuss  these various examples, let us consider the ``upgraded'' version of the Lense--Thiring spacetime \eqref{LTHD}, which possesses many interesting geometric properties.

\section{Improved Lense--Thirring spacetimes}

 Recently, there has been increasing interest in constructing certain variants of the Lense--Thirring metric~\cite{Baines:2020unr,Baines:2021qaw,Gray:2021toe,Baines:2021qfm}. These variants are constructed on the principle of including $\mathcal{O}(a^2)$ terms in the metric to achieve a form of `improved' behaviour. For example, the variants have finite Kretschmann scalar on the horizon, and can be cast into the manifestly regular  Painlev{\'e}--Gullstrand form~\cite{Martel:2000rn, Faraoni:2020ehi}. The inclusion of these terms, therefore, justifies the interpretation of these metrics as slowly rotating black holes. However, the $\mathcal{O}(a^2)$ terms \textit{do not} arise from solving the field equations at $\mathcal{O}(a^2)$, and the resulting metrics, viewed as physical solutions, still represent $\mathcal{O}(a)$ solutions to the field equations of the relevant theory. Taken as `off-shell' metrics in their own right, the variants provide physically-motivated examples of geometries with certain desirable properties,  the most remarkable one being that they admit a  tower of exact Killing tensors  that  rapidly grows with the number of spacetime dimensions~\cite{Gray:2021toe}. 

Thus far, variants of the Lense--Thirring spacetime considered in the literature have focused on vacuum Einstein gravity (see, however, \cite{Gray:2021toe} for including simple matter). As discussed in the introduction, this case has a number of very special properties, and there is no good reason to expect those properties to hold in general. Yet somewhat surprisingly,  the same behaviour extends to a much more general family of Lense--Thirring metrics. In what follows we are going to consider the following variant of the Lense--Thirring ansatz \eqref{LTHD} for  multiply-spinning black holes: 
\be\label{LTHDimproved}
ds^2=-Nfdt^2+\frac{dr^2}{f}+r^2\sum_{i=1}^m \mu_i^2\Bigl(d\phi_i+\frac{\sum_{j=1}^mp_{ij}a_j}{r^2} dt\Bigr)^2+r^2(\sum_{i=1}^m\!d\mu_i^2)+\epsilon r^2 d\nu^2\; .
\ee
 Obviously, the metric possesses the same `freedom' as the  metric \eqref{LTHD}. 
The functions $f, N, p_{ij}$ are functions of the radial coordinate $r$, and the 
coordinates $\mu_i$ are $\nu$ obey the constraint \eqref{constr}. Formally, it can be obtained by `completing the $t-\phi$ square' in \eqref{LTHD};  an alternative motivation is provided in App.~\ref{AppA}. In App,~\ref{AppB} we provide an orthonormal frame for the metric.  As we shall see below, this `tiny modification' has far reaching consequences for the properties of the corresponding spacetime.

First, the metric admits a Killing horizon generated by the following Killing vector:
\be
\xi=\partial_t+\sum_{i=1}^m \Omega_i \partial_{\phi_i}\,,
\quad 
\Omega_i= -\sum_{j=1}^m\frac{p_{ij}a_j}{r^2}\Bigr|_{r=r_+}\,,
\ee
where $r_+$ is the location of the horizon -- the largest root of $f(r_+)=0$. 
The horizon is surrounded by the ergoregion, inside of which the  Killing vector $\partial_t$ has negative norm. Due to this ergoregion, the metric will exhibit superradiant phenomena  \cite{Brito:2015oca}.

Second, the metric is regular on the horizon, and near its vicinity admits the Painlev{\'e}--Gullstrand form. Under the following coordinate transformation:
\ba
dt&=&dT-\sqrt{\frac{1-f}{N}}\frac{dr}{f}\,, \nonumber\\
d\phi_i&=& d\Phi_i+\frac{\sum_j p_{ij}a_j}{r^2}
\sqrt{\frac{1-f}{N}}\frac{dr}{f}\,,
\ea
 we recover
\ba\label{LTHD2}
ds^2&=&-NdT^2+\Bigl(dr+\sqrt{N(1-f)}dT\Bigr)^2+r^2\sum_{i=1}^m \mu_i^2\Bigl(d\Phi_i+\frac{\sum_j p_{ij}a_j}{r^2} dT\Bigr)^2\nonumber\\
&& \qquad+r^2(\sum_{i=1}^m\!d\mu_i^2)+\epsilon r^2d\nu^2\,,
\ea
which is manifestly regular on the horizon, 
and  the $T=const$. slices are manifestly flat.\footnote{If the $\mathcal{O}(a^2)$ corrections are considered to have physical effects, then it should be noted that the stress tensor associated with them fails to satisfy the classical energy conditions. Of course, this does not affect the results below concerning the hidden symmetry structure of the metrics, and moreover the falloff of the $\mathcal{O}(a^2)$ terms is sufficiently fast that inclusion of classical matter, e.g. an electromagnetic field, restores the energy conditions. }

Most importantly, the metric \eqref{LTHD2} also admits a rapidly growing tower of Killing tensors which guarantee separability of the Hamilton-Jacobi equations for geodesics and Klein-Gordon equation for scalar fields -- see App.~\ref{AppB}.  Generalizing \cite{Gray:2021toe}, these can be generated as follows. Define the set $S=\{1,..,m\}$ and let $I\in P(S)$ where $P(S)$ is the power set of $S$. Then we may define the  following objects:
\begin{align}
	2b^{(I)}&\equiv r^2(dt+\sum_{i\in I}a_i\mu_i^2d\phi_i)\,,\quad\,h^{(I)}\equiv db^{(I)}\,,\\
	f^{(I)}&\equiv \frac{\sqrt{N}}{(|I|+1)!}*\big(\underbrace{h^{(I)}\wedge \dots \wedge h^{(I)}}_{|I|+1\ \mbox{\tiny times}}\big)\,,
\end{align}
where $|I|$ denotes the size of the set $I$. These generate the following exact rank-2 Killing  tensors:
\begin{align}
K^{(I)}_{ab}&=\big(\prod_{i\in I
.
} a_i\big)^{-2}\,(f^{(I)}\cdot f^{(I)})_{ab}\,,\quad 
 K^{(I)}_{(ab;c)}=0\,,
\end{align}
where we  have defined 
\be 
(\omega_1 \cdot \omega_2)_{ab}=\frac{1}{p!}\omega_{ac_1\dots c_p}\omega_b{}^{c_1\dots c_p}
\ee 
for any $(p+1)$-forms $\omega_1,\omega_2$. Explicitly, these Killing tensors can be written as:
\begin{align}\label{KTs}
K^{(I)}=&\sum\limits_{i\not\in I}^{m-1+\epsilon}\bigg[\bigr(1-\mu_i^2-\!\sum_{j\in I}\mu_j^2\bigr)(\partial_{\mu_i})^2 -2\!\!\sum_{j\not\in I\cup\{i\}}\!\! \mu_i\mu_j\,\partial_{\mu_i}\partial_{\mu_j} \bigg]  +\sum\limits_{i\not\in I}^{m}\bigg[\frac{1-\sum_{j\in I}\mu^2_j}{\mu_i^2}(\partial_{\phi_i})^2\bigg] \,.
\end{align}
See App.~\ref{AppB} for details of the Killing tensors in the orthonormal frame. Now, since 
$\sum_{i=0}^{m-3}{m \choose i}$ of these are reducible, we have in total 
\be
k=\sum_{i=0}^{m-2+\epsilon}{m \choose i}-\sum_{i=0}^{m-3} {m \choose i} =\frac{1}{2} m (m-1+2\epsilon)
\ee
irreducible rank-2 Killing tensors in $d$ dimensions. 
 Note that this tower increases quadratically with the number of dimensions, contrary to the tower of rank-2 Killing tensors in exact Kerr--NUT--AdS spacetimes \cite{Frolov:2017kze}, which only grows linearly with $d$.

 Moreover, further non-trivial Killing tensors of higher rank,
\be
\nabla_{(a} K_{b_1\dots b_p)}=0\,,  
\ee
 can be generated via Schouten--Nijenhuis brackets, which for symmetric  
$A^{a_1\dots a_p}$ and $B^{b_1\dots b_q}$, are defined as \cite{schouten1940uber,nijenhuis1955jacobi}
\ba\label{SN}
[A,B]_{\mbox{\tiny SN}}^{a_1\dots a_{p+q-1}}&=&
pA^{c (a_1\dots a_{p-1}}\nabla_c B^{a_p\dots a_{p+q-1})} 
 -qB^{c(a_1\dots a_{q-1}}\nabla_c A^{a_q \dots a_{q+p-1})}\,. 
\ea
For a sufficiently high number of dimensions,  greater than six, there are at least two Killing tensors for which \eqref{SN} does not vanish. The generalized Lense--Thirring metrics \eqref{LTHDimproved} thus provide an example of physically motivated spacetimes with a tower of irreducible higher-rank Killing tensors.   Whether or not this tower ultimately terminates due to closure of the bracket in \eqref{SN} is currently unknown. For previous constructions of  spacetimes with higher rank Killing tensors, see \cite{Brink:2008xy, Gibbons:2011hg, Gibbons:2011nt, Cariglia:2015fva}.

Let us finally stress that  all the above hidden symmetries exist regardless of the form of the functions $p_{ij}=p_{ij}(r)$ and $N=N(r)$. In particular, when all $p_{ij}$ vanish, 
\be
p_{ij}=0\,, 
\ee
or alternatively, when all $a_i$ are zero, we recover the static spherically symmetric metric and all of the above Killing tensors become reducible -- given as sums of products of subsets of rotational Killing vectors.

\section{Vacuum solutions in higher curvature gravities}\label{Sec:vacuum}
In this section, rather then constructing the generalized Lense--Thirring solutions for concrete higher curvature gravities, we concentrate on answering the question: are there theories where the generalized Lense--Thirring solutions \eqref{LTHD} (or equivalently \eqref{LTHDimproved}) take a simple form \eqref{pi_Einstein}. That is, are they fully characterized by a single metric function of the associated static metric?

\subsection{Multiply-spinning Lovelock solutions}

An example of a higher curvature gravity where the answer is (perhaps remarkably) yes and \eqref{pi_Einstein} holds (even in the  multiply-spinning case), is  Lovelock gravity~\cite{Lovelock:1971yv}.  
This is a special higher curvature theory constructed in such a way so that the corresponding  equations of motion 
remain  second-order for any metric.  In $d$ spacetime dimensions, the Lovelock Lagrangian density reads
\be
\mathcal{L}_{\rm L} = \frac{1}{16 \pi } \bigg(\frac{(d-1)(d-2)}{\ell^2} + R 
+ \sum_{n=2}^{\lfloor (d-1)/2 \rfloor} \lambda_n \frac{(d+1-2n)!}{(d-1)!}(-1)^n \ell^{2n-2} \mathcal{X}_{2n} \bigg)\,,
\ee
where the Euler densities $\mathcal{X}_{2n}$ are given by
\be 
\mathcal{X}_{2n} = \frac{1}{2^n} \delta_{\nu_1 \dots \nu_{2n}}^{\mu_1 \dots \mu_{2n}} R_{\mu_1 \mu_2}^{\nu_1 \nu_2} \cdots R_{\mu_{2n-1} \mu_{2n}}^{\nu_{2n-1}\nu_{2n}} \, ,
\ee
$\delta_{\nu_1 \dots \nu_{2n}}^{\mu_1 \dots \mu_{2n}}$ are the generalized (antisymmetric) Kronecker delta symbols,  
and $\lambda_n$ are arbitrary dimensionless couplings\footnote{In the asymptotically flat case, the powers of $\ell$ appearing in the action can be absorbed into the couplings, making them dimensionful.}.

A particular example  is the Gauss--Bonnet theory ($k=2$), for which the series truncates at quadratic order in the curvature, and whose equations of motion become non-trivial in $d\geq 5$ dimensions. Likewise, $k$th-order Lovelock gravity is non-trivial in $d\geq (2k+1)$ number of spacetime dimensions.

The Lense--Thirring metric was first considered in the context of Gauss--Bonnet theory in~\cite{Kim:2007iw} for the case of a single rotation parameter in arbitrary dimensions, and for third-order Lovelock theory in~\cite{Yue:2011et}. The general case, for a single rotation parameter, was considered in~\cite{Camanho:2013pzg} -- see also the appendix of~\cite{Adair:2020vso}. These investigations found that in vacuum Lovelock gravity the Lense--Thirring metric takes the form of Eq.~\eqref{LTHDein}, that is, it is characterized by the static metric function.

Our main observation is that this property extends to all Lovelock theories and with multiple rotation parameters. In other words, the general multi-spinning Lense--Thirring metric to arbitrary order Lovelock theory  takes the form 
\eqref{LTHDimproved} with \eqref{pi_Einstein} holding, where 
the static metric function $f$ is determined in the usual way as a solution of the Wheeler polynomial,
\be
 \frac{16 \pi M \ell^2}{(d-2) \Omega_{d-2} r^{d-1}}  = 1 - \psi + \sum_{n=2}^{ \lfloor (d-1)/2 \rfloor} \lambda_n \psi^n \, ,\quad \psi \equiv \frac{\ell^2(f-1)}{r^2} \, .
\ee
This suggests that  the Newmann--Janis trick, or its higher-dimensional analogue,  can be used to generate slowly rotating vacuum solutions --- valid to linear order in the rotation parameter $a$ --- in Lovelock gravity.

\subsection{A family of generalized quasi-topological gravities}
Let us next consider a general higher curvature theory that includes up to quartic powers of the Riemann curvature \cite{Ahmed:2017jod}, with the corresponding 
Lagrangian density written as 
\be\label{Lgeneral} 
\mathcal{L} =\frac{1}{16 \pi}\Bigl(\frac{(d-1)(d-2)}{\ell^2}+ R + \sum_{i} \alpha_i \mathcal{R}_i^{(2)} + \sum_{i} \beta_i \mathcal{R}_i^{(3)} + \sum_{i} \gamma_i \mathcal{R}_i^{(4)}\Bigr) \, ,
\ee
where $\{\alpha_i, \beta_i,\gamma_i \}$ are (now dimensionful) couplings, the quadratic terms are
\be
\mathcal{R}^{(2)}_1 = R_{abcd} R^{abcd} \, , \quad \mathcal{R}^{(2)}_2 = R_{ab} R^{ab} \, , 
\quad
\mathcal{R}^{(2)}_3 = R^2 \, ,
\ee
the cubic terms are,
\begin{align}
\mathcal{R}^{(3)}_1 &= R_{a}{}{^c}_{b}{}^{d}R_c{}^e{}_d{}^f R_e{}^a{}_f{}^b \, ,
\quad
\mathcal{R}^{(3)}_2 = R_{ab}{}^{cd} R_{cd}{}^{ef} R_{ef}{}^{ab} \, ,\quad \mathcal{R}^{(3)}_3 = R_{abcd}R^{abc}{}_e R^{de} \,,
\nonumber\\
\mathcal{R}^{(3)}_4 &= R_{abcd}R^{abcd}R \,,\quad
\mathcal{R}^{(3)}_5 =  R_{abcd}R^{ac}R^{bd}\, ,
\quad
\mathcal{R}^{(3)}_6 = R_a{}^b R_b{}^c R_c{}^a \, ,
\nn\\
\mathcal{R}^{(3)}_7 &= R_a{}^b R_b{}^a R \, ,
\quad
\mathcal{R}^{(3)}_8 = R^3 \, ,
\end{align}
and the 26 independent quartic terms are listed in App.~\ref{AppC}.  Below, we will constrain the couplings so that the theories in question have certain desirable properties.

The field equations for this general theory can be written in the relatively compact form
\be\label{EOM} 
\mathcal{E}_{ab} = P_{a}{}^{cde}R_{bcde} - \frac{1}{2} g_{ab} \mathcal{L} - 2 \nabla^c \nabla^d P_{a c d b} = 0\,,
\ee
where the (somewhat cumbersome) expressions 
\be 
P^{abcd} \equiv \frac{\partial \mathcal{L}}{\partial R_{abcd}}  
\ee
for $P_{abcd}$ can be found in App.~\ref{AppC}.

In what follows we shall seek to determine which of the above theories admits the generalized Lense--Thirring solution of the form 
\eqref{pi_Einstein}. In particular, this means that we require  
\be\label{N1}
N=1\,,
\ee
and need to restrict to theories whose static solutions have the same property. The reason for this is that at leading order in $a$, the $\mathcal{E}_{tt}$ and $\mathcal{E}_{rr}$ components of the field equations are unmodified. Therefore, having $N=1$ in the Lense--Thirring metric is equivalent to having $N=1$ in the static solution. 
The family of theories for which this holds has been classified, and they are known as \textit{generalized quasi-topological (GQT) gravities}~\cite{Hennigar:2017ego, Bueno:2017sui}. Special noteworthy cases include Lovelock gravity~\cite{Lovelock:1971yv} and quasi-topological gravity~\cite{Oliva:2010eb, Myers:2010ru}, but the GQT class (which includes  \eqref{Lgeneral})
is much broader than this.

GQTs have been quite extensively studied and possess a number of appealing properties.  The most general kinds  have formulations in
4 spacetime dimensions, unlike the Lovelock subclass.
While all but the Lovelock family of theories possess higher-order equations of motion on general backgrounds, in all cases the equations are integrable and reduce to at most second-order when restricted to static and spherically symmetric backgrounds.  Interestingly, this property extends also to other simple metrics including Taub-NUT/Bolt solutions~\cite{Bueno:2018uoy}, cosmological solutions~\cite{Arciniega:2018fxj, Arciniega:2018tnn}, near horizon geometries~\cite{Cano:2019ozf}, and the four-dimensional Lense--Thirring metric~\cite{Adair:2020vso}. This feature, taken in tandem with the fact that GQT gravities provide a basis for an effective field theory of vacuum gravity~\cite{Bueno:2019ltp, Bueno:2019ycr}, makes the members of this family  physically relevant gravitational theories amenable to explicit calculation. Beyond this, these theories have been useful gravitational toy models. For example, in the context of the AdS/CFT correspondence, they have be utilized to generate a series of conjectures for universal properties of CFTs~\cite{Bueno:2018yzo, Bueno:2020odt}. 

That a theory belongs to the GQT class imposes certain constraints on the couplings. In brief,  the only quadratic theory satisfying these conditions is the Gauss--Bonnet density, corresponding to
\be 
\alpha_1 = \alpha \, , \quad \alpha_2 = - 4\alpha \, , \quad \alpha_3 = \alpha \, .
\ee
The cubic theories satisfying these conditions were classified in~\cite{Hennigar:2017ego}. In four dimensions, there is a four-parameter family of theories that meet this condition, while in higher dimensions, there is a three-parameter family -- c.f. Eqs.~(2.6) and (2.13) of ~\cite{Hennigar:2017ego}. Furthermore, the quartic theories satisfying these conditions were classified in~\cite{Ahmed:2017jod}. Again, there is a distinction between four and higher dimensions: there is a 19-parameter family of theories in four-dimensions, but a 17-parameter family of theories in higher dimensions --- c.f. appendix A and Eq.~(2.24) of~\cite{Ahmed:2017jod}. These constraints will be imposed in all calculations  in the rest of this section. 

The second requirement in \eqref{pi_Einstein}, namely\footnote{For brevity, we write $p_i$ whenever the metric functions are assumed to couple to only one rotation parameter, i.e., $p_{ij}=p_i\delta_{ij}$}
 \be\label{f-1}
p_{i}=(f-1)\,, 
\ee
imposes further restrictions. Inserting this ansatz into the equations of motion \eqref{EOM}, we find that a \textit{sufficient} condition for their resultant consistency reads\footnote{Note that this condition actually holds independently for each term appearing in the Lagrangian.}
\be\label{LLcon}
\Bigl(\mathcal{E}_t^{\phi_i} - \frac{1}{(d-2) r} \frac{d \mathcal{E}_t^t}{dr}\Bigr)\Bigr|_{p_{i}=(f-1)}=0 \,.
\ee
This condition is, for example, satisfied by the Einstein and Lovelock theories. 
However, it is not clear to us whether this condition is also necessary, or whether there might exist other  theories with \eqref{f-1} that violate \eqref{LLcon}.

In the following we shall try to construct a theory in four and five dimensions that admits  solutions of the form \eqref{pi_Einstein}.  In particular, we will be interested in {\it nontrivial} theories with this property. By this we mean theories that lead to genuine corrections to the metric, rather than simply admitting the Einstein gravity solution as one possible solution. There are a few different kinds of theories that we lump into the term `trivial'. For example:
\begin{enumerate}
\item There are dimension-dependent identities for the Riemann tensor that force certain terms to vanish identically for all metrics below a critical dimension. For example, the cubic Lovelock density vanishes identically in five and lower dimensions.
\item There are certain terms that are topological invariants and so make no dynamical contributions to the field equations in certain dimensions. For example, this is the case for the $k^{th}$ Lovelock lagrangian in $d = 2k$.
\item There are certain combinations of combinations of curvature tensors that, while not identically zero for all metrics, may make no contribution to a certain class of metrics, but are not topological invariants. 
\item Theories that are constructed solely from the Ricci curvature. Such theories, in the asymptotically flat case, will admit the Einstein gravity solution as one particular solution to the equations of motion.
\end{enumerate}  
With this clarification in mind, we now consider the cases of four and five dimensions separately.

\subsection{Uniqueness of Einstein gravity in four dimensions}

In this subsection we show that among all the GQT theories \eqref{Lgeneral}, Einstein gravity is the only nontrivial theory in four dimensions that admits \eqref{pi_Einstein}.  The hint that this might be the case follows from the condition \eqref{LLcon}. By considering the most general action up to quartic order and imposing \eqref{LLcon}, we find that no non-trivial theory remains. However, since \eqref{LLcon} might be only sufficient but not necessary for \eqref{pi_Einstein} to happen, we are not done yet.

To complete the proof, we thus adopt a  brute-force approach. Namely, to see if a $p = f - 1$ theory exists, rather than constructing full solutions to the (approximate) equations of motion, we focus on constructing asymptotic solutions, valid at large $r$.  As we shall see, the requirement \eqref{pi_Einstein} is strong enough to eliminate all GQT theories except Einstein gravity. Of course, our asymptotic solutions can be extended to full solutions in these theories -- we leave a full analysis of this for future work\footnote{See, for example, \cite{Adair:2020vso} for a full analysis in the particular case of Einsteinian Cubic Gravity.}.

We take the following ansatz for the metric functions (setting for simplicity $\Lambda=0)$:  
\be 
f = 1 - \frac{2M}{r} + \sum_{i = 2} \frac{b_i}{r^i} \, , \quad p = - \frac{2 M}{r} + \sum_{i=2} \frac{c_i}{r^i} \, .
\ee
We then insert these into the field equations (after constraining the couplings to be of the GQT class), and solve order-by-order.  While it is easy enough to work to arbitrarily high order, here we present the solution up to the order necessary to see the main result:
\begin{align}
f  =& 1 - \frac{2 M}{r} + \frac{108 M^2 \delta_1}{7 r^6} - \frac{184 M^3 \delta_1}{7 r^7} + \frac{432 M^3 \zeta_1}{5 r^9} - \frac{776 M^4 \zeta_1}{5 r^{10}} - \frac{15552 M^3 \delta_1^2}{7 r^{11}} + \frac{443232 M^4 \delta_1^2}{49 r^{12}}\nn\\ 
&- \frac{64032 M^5 \delta_1^2}{7 r^{13}} 
- \frac{1430784 M^4 \delta_1 \zeta_1}{35 r^{14}} + \frac{1157760 M^5 \delta_1 \zeta_1}{7 r^{15}}\nn\\
& - \frac{64 M^4 \delta_1 (4467379 M^2 \zeta_1 - 38053800 \delta_1^2 - 61236000 \delta_1 \delta_2 )}{1715 r^{16}} 
\nn\\
&- \frac{31104 M^5(2706525 \delta_1^3 + 4681250 \delta_1^2 \delta_2 + 46991 \zeta_1^2)}{8575 r^{17}}\nn\\
& + \frac{1728 M^6 (111394125\delta_1^3 + 20648250 \delta_1^2 \delta_2 +  3411478 \zeta_1^2)}{8575 r^{18}} 
+ \mathcal{O}(r^{-19}) \, ,
\\
p =& - \frac{2M}{r} - \frac{184 M^3 \delta_1}{7 r^7} + \frac{1728 M^3 ( \zeta_1 - \zeta_2)}{11 r^{9}} - \frac{776 M^4 \zeta_1}{5 r^{10}} + \frac{77760 M^3 \delta_1^2}{637 r^{11}} + \frac{198288 M^4 \delta_1^2}{49 r^{12}}\nn\\
& - \frac{64032 M^5 \delta_1^2}{7 r^{13}} 
- \frac{2592 M^4 \delta_1 (976\zeta_1 - 965\zeta_2 + 100\zeta_3)}{35 r^{14}} \nn\\
&+ \frac{5184 M^5 \delta_1 (26751\zeta_1 - 17435\zeta_2 + 1700\zeta_3)}{595 r^{15}} 
- \frac{64 M^4 \delta_1 ( 4467379 M^2 \zeta_1 + 2199150\delta_1^2)}{1715 r^{16}}
\nn\\
&-\frac {5184{M}^{5}}{162925 r^{17}}\Bigl( 75690675\delta_{{1}}^{3}+127575000
\delta_{{1}}^{2}\delta_{{2}}+4153044\zeta_{{1}}^{2}+10183670\zeta_{{2}}\zeta_{{1}}\nn\\
&\qquad\qquad-2346120\zeta_{{3}}\zeta_{{1}}-13514200\zeta_{{2}}^{2}+
5213600\zeta_{{3}}\zeta_{{2}} \Bigr)
\nn\\
&+\frac{1728{M}^{6}}{1225 r^{18}}\Bigl( 9393975\delta_{{1}}^{3}+16380000
\delta_{{1}}^{2}\delta_{{2}}+428848\zeta_{{1}}^{2}+490784\zeta_{{2}}\zeta_{{1}}-105840\zeta_{{3}}\zeta_{{1}}\nn\\
&\qquad \qquad -646800\zeta_{{2}}^{2}+235200
\zeta_{{3}}\zeta_{{2}} \Bigr)
 + \mathcal{O}(r^{-19}) \, ,
\end{align}
where the couplings enter into the solutions through the combinations
\begin{align}
\delta_1 &= \beta_1 + 2 \beta_2 \, , \quad \delta_2 = \beta_2 - \frac{\beta_6}{2} \, , \nn
\\
\zeta_1 &= \gamma_1 + \frac{5 \gamma_2}{2} + 2 \gamma_4 + 4 \gamma_5 + 4 \gamma_6 + 16 \gamma_7 \, ,
\nn\\
\zeta_2 &= \gamma_2 + 4 \gamma_5 + 2 \gamma_6 + 8 \gamma_7 \, ,
\nn\\
\zeta_3 &= \frac{\gamma_3}{2} + \gamma_4  + 13 \gamma_5 + \frac{\gamma_6}{2} - 18 \gamma_7  + \frac{\gamma_8}{4} + 3 \gamma_9 + \frac{5 \gamma_{10}}{2} + \gamma_{13} + \frac{5 \gamma_{14}}{4} + \frac{\gamma_{15}}{2} \, ,
\nn\\
\zeta_4 &= \frac{18354}{4355} \left(-8 \gamma_5 + 16 \gamma_7 - 2 \gamma_9 - 2 \gamma_{10} - \gamma_{13} - \gamma_{14} \right) \, .
\end{align}
If we then demand that $p = f-1$ it becomes clear from the above that the combinations of couplings entering the metric function must vanish: $\delta_i = \zeta_i = 0$. We have shown only the first 18 terms in the above, as this is the minimal number to see that all the quartic contributions must vanish.\footnote{Note that $\zeta_4$ only enters at $O(r^{-19})$ but does not modify the metric function $p(r)$ if the other combinations of couplings vanish, i.e. $\delta_i = \zeta_i = 0$ for $i=1,2,3$.} We have carried out the expansion to much higher order in Maple, obtaining the same result. 

Note that this does not force all the individual couplings to be zero, just the combinations given by $\delta_i$ and $\zeta_i$. The reason for that is simple: On one hand, several densities contribute in the same way to the field equations, while on the other hand, certain combinations of densities are trivial. For example, the cubic and quartic Lovelock Lagrangian densities vanish identically in four dimensions, which provides a non-trivial relationship between the densities. This, along with other such relationships, is why there can be non-trivial couplings in the action, while the solution itself receives no corrections.

The conclusion of the above analysis is simple: In four dimensions, and up to quartic order in curvature, Einstein gravity is the only non-trivial theory for which the Lense--Thirring metric takes the form~\eqref{pi_Einstein}. Note that, while we have worked with $\Lambda = 0$ here, the same conclusions go through when a cosmological constant is included.

\subsection{Five dimensions: special GQT theories}

 As we shall now show, in five dimensions, there exist special GQT theories, e.g. \eqref{L4} below, which admit generalized Lense--Thirring solutions of the form \eqref{pi_Einstein}.

In five dimensions, the metric is more complicated, as it is possible to have two independent rotation parameters. We begin by exploring the possibility of $p_i = (1 - f)$ metrics by attempting to force the relationship~\eqref{LLcon}. Unlike the four-dimensional case, here we find success. Upon imposing the following constraints on the cubic couplings (in addition to the constraints mentioned above that enforce that the theory is a member of the GQT class of theories):
\begin{align}
\beta_1 &= -2 \beta_2 \, , \quad \beta_3 = - 6 \beta_2 \, ,\nn
\\ 
\gamma_{3} &= \frac{71 \gamma_2}{13} + \frac{666 \gamma_4}{13} - \frac{2124 \gamma_5}{13} - \frac{536 \gamma_6}{13} - 216 \gamma_7 + 33 \gamma_1\, ,
\nn\\
\gamma_8 &= - \frac{108 \gamma_2}{13} - \frac{1016 \gamma_4}{13} + \frac{2768 \gamma_5}{13} + \frac{704 \gamma_6}{13} + 288 \gamma_7 - 48 \gamma_1 \, , 
\nn\\
\gamma_9 &= 0 \, ,
\\
\gamma_{10} &= -\frac{373 \gamma_2}{65} - \frac{804 \gamma_4}{65} + \frac{456 \gamma_5}{65} - \frac{288 \gamma_6}{65} - \frac{32 \gamma_1}{5} \, ,
\nn\\
\gamma_{15} &= -11 \gamma_{13} - \gamma_{16} + \frac{13071 \gamma_2}{130} + \frac{15764 \gamma_4}{65} - \frac{15856 \gamma_6}{65} - 24 \gamma_7 - \frac{25 \gamma_{14}}{2} + \frac{642 \gamma_1}{5} \, , 
\nn\\
\gamma_{18} &= 3 \gamma_{11} + 4 \gamma_{13} - \frac{11137 \gamma_2}{260} - \frac{7949 \gamma_4}{65} + \frac{10306 \gamma_5}{65} - \frac{358 \gamma_6}{65} + 96 \gamma_7 + 5 \gamma_{14} - \frac{659 \gamma_1}{10} \, ,\nn
\end{align}
any non-trivial contributions are completely eliminated at that order. 

However, there persists a five-parameter family of GQT theories at quartic order that are non-trivial. All five members of this family make exactly the same contributions to the field equations of the Lense--Thirring ansatz. A particular example of this family of theories, corresponding to setting $\gamma_6 = 614 \gamma_1 /5$ and all other couplings besides $\gamma_1$ to zero, is
\begin{align}\label{L4}
\mathcal{L}^{(4)} =& \frac{68}{5}  R_{a}{}^{c} R^{ab} R_{b}{}^{d} R_{cd} + \frac{2797}{90}  R_{ab} R^{ab} R_{cd} R^{cd} + \frac{1}{5}  R_{a}{}^{c} R^{ab} R_{bc} R -  \frac{721}{45}  R_{ab} R^{ab} R^2 \nn\\
&+ \frac{1306}{15}  R^{ab} R^{cd} R R_{acbd} 
+ \frac{433}{36}  R^2 R_{abcd} R^{abcd} -  \frac{659}{10}  R^{ab} R R_{a}{}^{cde} R_{bcde} \\
&-  \frac{2764}{15}  R_{a}{}^{c} R^{ab} R^{de} R_{bdce} + \frac{28}{5}  R^{ab} R^{cd} R_{ac}{}^{ef} R_{bdef} + \frac{31}{10}  R R_{ab}{}^{ef} R^{abcd} R_{cdef}
\nn\\
& + \frac{614}{5}  R_{a}{}^{c} R^{ab} R_{b}{}^{def} R_{cdef} - 48  R^{ab} R_{a}{}^{cde} R_{b}{}^{f}{}_{d}{}^{h} R_{cfeh} -  \frac{1174}{45}  R_{ab} R^{ab} R_{cdef} R^{cdef} 
\nn\\
&-  \frac{32}{5}  R^{ab} R_{a}{}^{c}{}_{b}{}^{d} R_{c}{}^{efh} R_{defh} +  R_{a}{}^{e}{}_{c}{}^{f} R^{abcd} R_{b}{}^{h}{}_{e}{}^{i} R_{dhfi} + 33  R_{ab}{}^{ef} R^{abcd} R_{c}{}^{h}{}_{e}{}^{i} R_{dhfi}\,.\nn
\end{align}
For these theories, the $\mathcal{E}_t^t$ field equation is third-order, but integrable and reduces to a second-order differential equation that can be solved for the metric function $f$. Upon substituting  $p_i = (f - 1)$, the $\mathcal{E}_t^{\phi_i}$ field equation satisfies the relationship~\eqref{LLcon} and takes the following form:\footnote{This is obtained after integrating the $\mathcal{E}_t^t$ component of the field equations, which requires an integrating factor $r^3$.} 
\be\label{quarticEQ} 
\frac{3 r^2(f-1)}{2} + \frac{9 \alpha (rf' - 2 f + 2)^2}{4 r^2} \bigg[r^2 f f'' - \frac{r^2 f'^2}{4} - \left(f + \frac{1}{3} \right) r f' + f^2 - \frac{4 f}{3} + \frac{1}{3} \bigg] = -  \frac{3m}{2}  \, .
\ee
At leading order, asymptotically, the correction to the Einstein gravity metric function due to this density is given by
\be
f(r) = 1 - \frac{m}{r^2} + \frac{224 \alpha m^3}{r^{12}} - \frac{192 \alpha m^4}{r^{14}} + \mathcal{O}(r^{-22}) \, .
\ee
The methods for solving these equations in general are now well-understood~\cite{Adair:2020vso}. Although it is not the purpose of the present paper, it would be interesting to construct the full solution of \eqref{quarticEQ} numerically and study the properties of the Lense-Thirring metric, and its AdS generalization, in this theory.

\section{Generalized Lense--Thirring solutions with matter}\label{Sec:matter}
 
We next shall write down a couple of generalized Lense--Thirring solutions with matter fields, for simplicity focusing on solutions in Einstein gravity. Namely, we write down  
the (Maxwell) charged Kerr-AdS solutions (example 1), the slowly rotating Kerr--Sen solution (example 2), and the solution of the $d=5$ minimal gauged supergravity (example 3). The purpose of these examples is to illustrate the following facts. 
\begin{itemize}
\item Example 1 shows that even with matter fields, one may still find solutions in the form \eqref{pi_Einstein}. 
\item Example 2 provides an example of the generalized Lense--Thirring spacetime with non-trivial $N$. 
\item Example 3 shows the possibility of having distinct metric functions $p_{ij}$ (distinct also from $f-1$) and illustrates a very interesting effect of `mixing' of rotation parameters,  induced by the Chern--Simons coupling.   
\end{itemize}
The latter two examples are constructed by taking the slow rotation limit of the full exact black hole solutions.

\subsection{Example 1: Charged Kerr-AdS in all dimensions}

The general multiply-spinning slowly rotating Kerr-AdS spacetimes are given by \eqref{LTHDein} and \eqref{fLambda} -- they can be cast in the form \eqref{LTHDimproved} with the special property  \eqref{pi_Einstein}.   
Slightly more generally, we can consider (possibly strongly) charged slowly rotating solutions of Einstein--Maxwell theory:
\be
{\cal L}_M=\frac{1}{16\pi}\Bigl(R-F_{ab}F^{ab}+\frac{(d-1)(d-2)}{\ell^2}\Bigr)\,. 
\ee
Interestingly, the presence of the electromagnetic field in this case does not spoil the property \eqref{pi_Einstein}, and the solution takes the form 
\eqref{LTHDimproved}, with $N=1$ and 
\be
p_i=f-1=-\frac{m}{r^{d-3}}+\frac{q^2}{r^{2(d-3)}}+\frac{r^2}{\ell^2}\,, 
\ee 
where $m$ and $q$ are parameters related to mass and charge, respectively.
The metric is accompanied by the Maxwell field, $F=dA$, where the vector potential $A$ takes the following form:
\be
A=-\sqrt{ \frac{d-2}{2(d-3)}}\frac{ q}{r^{d-3}}\Bigl[dt-\sum_{i=1}^m \bigl(a_i \mu_i^2 d\phi_i-
\frac{a_i^2\mu_i^2 p_i}{r^2f}dr\bigr)\Bigr]\,, 
\ee
where the last term was introduced in order that the field invariant $F_{ab}F^{ab}$ be finite on the horizon, $f=0$.

\subsection{Example 2: Slowly rotating Kerr--Sen spacetime}

 Let us next consider the following 4-dimensional low-energy effective action describing heterotic string theory: 
\be\label{action}
{\cal L}= \frac{e^{\Phi}}{16\pi}\Bigl(R + g^{a b}\partial_{a}\Phi\partial_{b}\Phi-F_{ab}F^{ab}-\frac{1}{12}H_{a b c}H^{a b c}\Bigr)\,,
\ee
where $g_{ab}$ represents the metric in the string frame, $\Phi$ is the dilaton field, $F = dA$ is the
Maxwell field strength, and $H = dB-2A\wedge F$ is a 3-form defined in terms of the vector
potential $A$ and a 2-form potential $B$.

This theory admits an exact rotating black hole solution, known as the Kerr--Sen metric 
\cite{Sen:1992ua}, which in the  standard Boyer--Lindquist-type coordinates and the string frame reads~\cite{Sen:1992ua,Houri:2010fr, Wu:2020cgf}:
\begin{align}\label{Senmetric}
{d}s^2\!&=\!e^{-\Phi}\Bigl(-\frac{\Delta_b}{\rho_b^2}\bigl({d}t-a\sin^2\!\theta{d}\phi\bigr)^2+\frac{\rho_b^2}{\Delta_b}{d}r^2+\frac{\sin^2\!\theta}{\rho_b^2}\Bigl[a{d}t-(r^2+2\,b\,r+a^2){d}\phi\Bigr]^2+\rho_b^2 {d}\theta^2\Bigr)\,,\nonumber\\
{B}&=\frac{2abr}{\rho_b^2}\,\sin^2\theta {d}t\wedge {d}\phi\,,\quad
A\!=\!-\frac{Q\,r}{\rho_b^2}\bigl({d}t-a\,\sin^2\!\theta {d}\phi\bigr)\,,\quad
e^{-\Phi}= \frac{\rho^2}{\rho_b^2}\,,
\end{align}
where the metric functions are given by
\be
\rho^2=r^2+a^2\cos^2\theta\,,\quad   \rho_b^2=\rho^2+2br\,,\quad  \Delta_b=r^2-2(M-b)r+a^2\,,
\ee
and the 3-form $H$ reads
\be\label{HHH}
H=-\frac{2\,b\,a}{\rho_b^4}{d}t\wedge {d}\phi\wedge\Bigl[\bigl(r^2-a^2\cos^2\!\theta\bigr)\sin^2\!\theta{d}r-r\Delta_b\sin 2\theta d\theta\Bigr]
\ee
where $M$ is the mass of the black hole, $Q$   its charge, $a$ its rotation parameter, and $b=Q^2/(2M)$ is the twist parameter.

Taking the $\mathcal{O}(a)$ expansion, and completing the square we recover the following generalized Lense--Thirring solution with non-trivial $N$ and $p$: 
\ba
ds^2&=&-Nfdt^2+
\frac{dr^2}{f}+r^2\sin^2\!\theta
\Bigl(d\phi+\frac{ap}{r^2}dt\Bigr)^2+r^2d\theta^2\,,\nonumber\\
A&=&-\frac{Q}{r+2b}\bigl(dt-a\sin^2\!\theta d\phi\bigr)
\,,\nn\\
B&=&\frac{2ab}{r+2b}\sin^2\!\theta dt\wedge d\phi\,,\quad e^\Phi=1+\frac{2b}{r}\,,
\ea
where
\be
f=1-\frac{2(M-b)}{r}\,,\quad 
N=\Bigl(1+\frac{2b}{r}\Bigr)^{-2}\,,\quad
p=N\Bigl(f-1-\frac{2b}{r}\Bigr)\,.
\ee
The corresponding exact Killing tensor reads 
\be
K=\frac{1}{\sin^2\!\theta} (\partial_\phi)^2+(\partial_\theta)^2\,.
\ee

\subsection{Example 3: Solution of $d=5$ minimal gauged supergravity}

 Finally, let us consider the action for the $d=5$ minimal gauged supergravity:
\be
{\cal L}=\frac{1}{16\pi}\Bigl(*(R-2\Lambda)-\frac{1}{2}F\wedge *F+\frac{1}{3\sqrt{3}}F\wedge F\wedge A\Bigr)\,, 
\ee
where $F=dA$.
The general rotating charged black hole was constructed in \cite{Chong:2005hr} and is known as the Chong--Cveti{\v c}--L{\"u}--Pope solution. Contrary to the original paper, we write it in a coordinate system that rotates at infinity as follows:
\ba\label{SUGRAmetric}
ds^2&=& d\gamma^2 - \frac{2q \nu \omega}{\Sigma}+\frac{\sigma \omega^2}{\Sigma^2} + \frac{\Sigma dr^2}{\Delta}
+ \frac{\Sigma d\theta^2}{S}\,,\quad
\nn\\
A &=& \frac{\sqrt{3}q\omega}{\Sigma}\,,
\ea
where we have defined
\ba\label{nuomega2}
\nu&=&\frac{ab}{\ell^2}dt-b\sin^2\!\theta d\phi-a\cos^2\!\theta d\psi\,,\quad \omega=dt+\frac{a\sin^2\!\theta d\phi}{\Xi_a}+\frac{b\cos^2\!\theta d\psi}{\Xi_b}\,,\nn\\
d\gamma^2&=&
\frac{\sin^2\!\theta}{\Xi_a}\Bigl[(r^2+a^2)d\phi^2-\frac{2a}{\ell^2}(r^2+a^2)dtd\phi-\frac{dt^2}{\ell^2}\bigl(\rho^2-(r^2+a^2)\frac{a^2}{\ell^2}\bigr)\Bigr]\nonumber\\
&&+\frac{\cos^2\!\theta}{\Xi_b}\Bigl[(r^2+b^2)d\psi^2-\frac{2b}{\ell^2}(r^2+b^2)dtd\psi-\frac{dt^2}{\ell^2}\bigl(\rho^2-(r^2+b^2)\frac{b^2}{\ell^2}\bigr)\Bigr]\,,
\ea
and
\ba
S&=& \Xi_a\cos^2\theta +\Xi_b\sin^2\theta\,,\quad
\Delta = \frac{(r^2+a^2)(r^2+b^2)\rho^2/\ell^2+q^2+2abq}{r^2}-2m\,,
\nn\\
\Sigma &=& r^2+a^2\cos^2\theta + b^2\sin^2\theta\,,\quad \rho^2=r^2+\ell^2\,,
\nn\\
\Xi_a &=& 1-\frac{a^2}{\ell^2}, \quad \Xi_b = 1-\frac{b^2}{\ell^2}\,,
\quad 
\sigma = 2m\Sigma -q^2+\frac{2abq}{\ell^2}\Sigma\,.
\ea
The black hole rotates in two different directions, corresponding to the rotation parameters $a$ and $b$, while the parameter $q$ is related to the black hole charge, and $m$ to its mass. The spacetime admits a principal Killing--Yano tensor with torison \cite{Kubiznak:2009qi}, which generates an exact Killing tensor.

Taking the linear $\mathcal{O}(a)$ and $\mathcal{O}(b)$ limit, and completing the square we obtain the following generalized Lense--Thirring solution:
\ba
ds^2&=&-fdt^2+\frac{dr^2}{f}+r^2d\theta^2+r^2\sin^2\!\theta \Bigl(d\phi+\frac{ ap_{aa} +bp_{ab} }{r^2}dt\Bigr)^2+r^2\cos^2\!\theta \Bigl(d\psi+\frac{bp_{bb}+ap_{ba}}{r^2}dt\Bigr)^2\,,\nonumber\\
A&=&\frac{\sqrt{3}q}{r^2}\Bigl(dt-a\sin^2\!\theta d\phi-b \cos^2\!\theta d\psi+\frac{a [ap_{aa}+bp_{ab}]\sin^2\!\theta +b[b p_{bb}+ap_{ba}]\cos^2\!\theta }{r^2 f}dr\Bigr)\,,
\ea
where 
\be
f= 1-\frac{2m}{r^2}+\frac{q^2}{r^4}+\frac{r^2}{\ell^2}\,, \quad p_{aa}=f-1=p_{bb}\,,\quad p_{ab}=\frac{q}{r^2}=p_{ba}\,.
\ee 
This solution provides a unique example where the presence of `Chern--Simons charge' `mixes' the rotation parameters in the metric functions $p_{ab}$, which are distinct from each other and distinct from $(f-1)$. 
The spacetime admits the following Killing tensor 
\be
K=\frac{1}{\sin^2\!\theta}(\partial_\phi)^2+\frac{1}{\cos^2\!\theta}(\partial_\psi)^2+(\partial_\theta)^2\,, 
\ee
inherited from the approximate Killing--Yano tensor with torsion.

\section{Conclusions}

Our purpose in this work has been two-fold. We began by considering variants of the Lense--Thirring metric that have recently attracted attention in the literature for possessing a number of `improved properties', such as the existence of a tower of \textit{exact} Killing tensors that rapidly grows with the number of dimensions.  The metrics of this kind so far considered in the literature represent a restricted class, as they have been studied only in the context of vacuum Einstein gravity, where the entire solution is characterized by the Schwarzschild metric function $f = 1 - 2M/r$. We have demonstrated that the same `improved properties' hold for a much more general class of slowly rotating metrics, that includes the vacuum Einstein gravity metric as a special case.

By deforming Einstein gravity through the addition of higher curvature terms in the action of up to quartic order, we demonstrated that in four dimensions there are no higher curvature theories for which the Lense--Thirring metric receives corrections and is still characterized by the metric function of the static black hole. In other words, four-dimensional vacuum Einstein gravity is the only \textit{non-trivial} theory with this property. However, in higher dimensions we demonstrated that additional theories exist for which this holds true. Lense--Thirring metrics in Lovelock gravity and certain quartic generalized quasi-topological gravities both receive corrections and are characterized still in terms of the metric function of the static black hole. In showing this, we have demonstrated that the Newman--Janis trick can be used to generate slowly rotating solutions (valid to linear order in the rotation parameter) in these theories. Of course, this does not immediately extend to the construction of fully rotating solutions in these theories, but one may wonder if the construction of full rotating solutions is easier in these theories than in the general case.

Finally, we have considered slowly rotating approximations to a number of known exact-solution rotating black holes that include matter. In doing so, we have provided examples where each additional term in the generalized ansatz is non-trivial.

For future work, it would be interesting to construct the full generalized Lense--Thirring solutions in higher curvature gravities, discussed here only asymptotically in Sec.~\ref{Sec:vacuum}.   At the same time, 
it would be interesting to extend our considerations to other effective theories with matter fields -- for example, involving a metric and a scalar. In this context, it is known that the Horndeski theory corresponding to the four-dimensional limit of Gauss-Bonnet gravity \cite{Lu:2020iav, Hennigar:2020lsl, Fernandes:2020nbq} allows for a Lense--Thirring metric characterized by the static metric function~\cite{Charmousis:2021npl}. One may then wonder if this is the unique Horndeski theory with that property.

\section*{Acknowledgements}
F.G. acknowledges support from the Natural Sciences and Engineering Research Council of Canada (NSERC) via a Vanier Canada Graduate Scholarship. This work was supported by the Perimeter Institute for Theoretical Physics and by NSERC. Research at Perimeter Institute is supported in part by the Government of Canada through the Department of Innovation, Science and Economic Development Canada and by the Province of Ontario through the Ministry of Colleges and Universities. 
Perimeter Institute and the University of Waterloo are situated on the Haldimand Tract, land that was promised to the Haudenosaunee of the Six Nations of the Grand River, and is within the territory of the Neutral, Anishnawbe, and Haudenosaunee peoples.

\appendix
\section{Regular slow rotation expansion of Kerr}\label{AppA}

In this appendix we attempt to physically motivate the improved Lense--Thirring form of the metric \eqref{LTHDimproved}. To do this, we construct a slowly rotating variant of the Lense--Thirring solution starting from the Kerr metric, writing it in Kerr ingoing coordinates, expanding to linear order in rotation parameter, and returning back to the Boyer--Lindquist coordinates. This yields  a metric that is a vacuum solution of Einstein equations to linear order in rotation parameter $a$, and which is manifestly regular on the horizon  when taken `as is'.
Since the transformation between Kerr and Boyer--Lindquist coordinates involves the rotation parameter, this metric carries certain $\mathcal{O}(a^2)$ corrections, slightly distinct, however, from the improved Lense--Thirring solution \eqref{LTHDimproved}.

Let us start from the Kerr metric written in the standard Boyer--Lindquist coordinates:
\ba
ds^2&=&-\frac{\Delta}{\Sigma}(dt-a\sin^2\!\theta d\phi)^2+\frac{\Sigma}{\Delta}dr^2+\Sigma d\theta^2+\frac{\sin^2\!\theta}{\Sigma}\Bigl[adt-(r^2+a^2)d\phi\Bigr]^2\,, \nn\\
\Sigma&=&r^2+a^2\cos^2\!\theta\,,\quad \Delta=r^2+a^2-2mr\,, 
\ea
and perform the following coordinate transformation to Kerr coordinates $\{v,\chi, r, \theta\}$: 
\be
dv=dt+\frac{r^2+a^2}{\Delta}dr\,,\quad d\chi=d\phi+\frac{a}{\Delta}dr\,. 
\ee
By expanding the resultant metric to linear order in $a$ we obtain:
\ba
ds^2&=&-fdv^2+2dvdr-\frac{4Ma}{r}\sin^2\!\theta dvd\chi
-2a\sin^2\!\theta d\chi dr+r^2\sin^2\!\theta d\chi^2+r^2 d\theta^2\,,\nn\\
f&=&1-\frac{2M}{r}\,. 
\ea
Using now the inverse transform to linear order in $a$:
\be
dv=dt+\frac{dr}{f}\,,\quad d\chi=d\phi +\frac{a}{r^2f}dr\,, 
\ee
we thus recover the following metric: 
\ba
ds^2&=&-fdt^2+\frac{dr^2}{f}+
r^2d\theta^2+r^2\sin^2\!\theta \Bigl(d\phi+\frac{a(f-1)}{r^2}dt\Bigr)^2\nonumber\\
&&
-\frac{a^2\sin^2\!\theta}{r^2f^2}\Bigl(f(1-f)dt+dr\Bigr)^2\,. 
\ea
Note that the first line is the improved Lense--Thirring metric \eqref{LTHDimproved}, while dropping all $\mathcal{O}(a^2)$ terms gives the ordinary Lense-Thirring metric.

\section{Improved Lense--Thirring spacetimes: further considerations} \label{AppB}

\subsection{Orthonormal frame}
Consider the following generalized Lense--Thirring metric \eqref{LTHDimproved} with a slight change of notation: rather than denoting by $\nu$, the constrained coordinate in even dimensions, we use $\mu_{m+\epsilon}$, then the metric and constraint simplify to
\be\label{LTHDimprovedApp}
ds^2=-Nfdt^2+\frac{dr^2}{f}+r^2\sum_{i=1}^m \mu_i^2\Bigl(d\phi_i+\sum_{j=1}^m\frac{a_jp_{ij}}{r^2} dt\Bigr)^2+r^2\sum_{i=1}^{m+\epsilon}d\mu_i^2\;,
\ee
 and
\be
\sum_{i=1}^{m+\epsilon} \mu_i^2=1\,.
\ee
Then one can calculate the orthonormal vielbeins $e^A=e_a^A\,dx^a$ which satisfy
\be
g_{ab}=\eta_{AB} e^A_a e^B_b\,.
\ee 
We find
\begin{align}
e^{\hat{t}}&=\sqrt{Nf}dt\,,\quad e^{\hat{r}}=\frac{dr}{\sqrt{f}}\,,\quad e^{\hat{\phi}_i}=r\mu_i\left(d\phi_i +\sum_{j=1}^m\frac{a_jp_{ij}}{r^2} dt\right)\,,\nonumber\\
e^{\hat{\mu}_i}&=\frac{r}{\sqrt{\mu_{m+\epsilon}^2+\sum\limits_{j=1}^{i-1} \mu_j^2 }}\left(\sqrt{\mu_{m+\epsilon}^2+\sum_{k=1}^{i} \mu_k^2 }\,\,\,d\mu_i +\!\!\!\!\sum_{l>i}^{\,\,\,m-(1-\epsilon)} \frac{\mu_i\mu_ld\mu_l}{\sqrt{\mu_{m+\epsilon}^2+\sum_{k=1}^{i} \mu_k^2 }}\right)\,,
\end{align}
with the following inverse $e_A=e_A^a\partial_a$
\begin{align}
e_{\hat{t}}&=\frac{1}{\sqrt{Nf}}\left(\partial_t-\sum_{j=1}^m\frac{a_jp_{ij}}{r^2}\partial_{\phi_j}\right)\,,\quad e_{\hat{r}}=\sqrt{f}\partial_r\, ,\quad e_{\hat{\phi}_i}=\frac{1}{r\mu_i}\,\partial_{\phi_i}\,,\nonumber\\
e_{\hat{\mu}_i}&=\frac{\sqrt{\mu_{m+\epsilon}^2+\sum\limits_{j=1}^{i-1} \mu_j^2 }}{r}\left(\frac{\partial_{\mu_i}}{\sqrt{\mu_{m+\epsilon}^2+\sum_{k=1}^{i} \mu_k^2 }} -\left(\frac{\mu_i}{\mu_{m+\epsilon}^2+\sum_{k=1}^{i-1}\mu_k^2}\right)\,\sum_{l=1}^{i-1} \frac{\mu_l\partial_{\mu_l}}{\sqrt{\mu_{m+\epsilon}^2+\sum_{k=1}^{i} \mu_k^2 }}\right)\,.
\end{align}
\subsection{Killing Tensors}
The separability and integrability of the spacetime requires $m$  Killing tensors\footnote{One of these is, of course, the trivial Killing tensor, i.e. the metric.} in addition to the $1+(m+\epsilon)$ Killing vectors $\partial_t$ and $\partial_{\phi_j}$. Now, we have seen the the metric \eqref{LTHDimproved} has a fast growing (with number of dimensions) tower of exact Killing tensors. Explicitly, as a reminder let us recall the following: given the set $S=\{1,..,m\}$, let $I\in P(S)$ where $P(S)$ is the power set of $S$, then we have the exact rank-2 Killing  tensors c.f \eqref{KTs}:
\begin{align}
K^{(I)}&=\!\!\!\!\!\sum\limits_{i\not\in I}^{m-1+\epsilon}\!\bigg[\bigr(1-\mu_i^2-\!\sum_{j\in I}\mu_j^2\bigr)(\partial_{\mu_i})^2 -2\!\!\!\!\sum_{j\not\in I\cup\{i\}}\!\!\!\! \mu_i\mu_j\,\partial_{\mu_i}\partial_{\mu_j} \bigg] +\sum\limits_{i\not\in I}^{m}\bigg[\frac{1-\sum_{j\in I}\mu^2_j}{\mu_i^2}(\partial_{\phi_i})^2\bigg] \,.
\end{align}
Moreover, we have verified up to $d=13$ that the SN bracket \eqref{SN} of any two Killing tensors vanishes if the intersection of the two set labels equals one of the two. That is, if $I_1\cap I_2=I_1$ or if $I_1\cap I_2=I_2$,
\be\label{SNprop}
[K^{(I_1)}, K^{(I_2)}]_{\mbox{\tiny SN}}=0\,.
\ee

In particular, we find there is a subset of these Killing tensors which are diagonal in the orthonormal basis. Let us denote $\tilde{m}=m-(1-\epsilon)$, and then define $Q\subset P(S)$ by
\be
Q=\Big\{\emptyset, \{\tilde{m}\}, \{\tilde{m}, \tilde{m}-1\},\dots, \{\tilde{m},\tilde{m}-1,\tilde{m}-2,\dots,2 \} \Big\}\footnote{Note that there are exactly $m-1$ elements in Q -- one from each `level' of subsets in $P(S)\backslash S$.}\,.
\ee
Then for all $J\in Q$
\be
K^{(J)}=r^2\left(1-\sum_{j\in J}\mu_j^2\right)\left( \sum_{i\not\in J}^{\tilde{m}}e_{\hat{\mu}_i} e_{\hat{\mu}_i}+\sum_{i\not\in J}^{m}e_{\hat{\phi}_i} e_{\hat{\phi}_i}\right)\,.
\ee
Since these elements of $Q$ satisfy the following nesting property:
\be
\emptyset \subset \{\tilde{m}\} \subset\{\tilde{m}, \tilde{m}-1\}\subset \dots \subset \{\tilde{m},\tilde{m}-1,\tilde{m}-2,\dots,2 \}\,,
\ee
these Killing tensors all mutually Schouten--Nijenhuis (SN) commute by \eqref{SNprop} guaranteeing the separability of the Hamilton--Jacobi equation~\cite{Frolov:2017kze}. 

Moreover, as these are diagonal in the orthonormal basis they commute (as matrices) with the Ricci tensor, and so, satisfy Carter's criterion~\cite{Carter:1977pq} $\nabla_a(k_\gamma{}^{[\alpha}R^{\beta]\gamma})=0$. Hence they define commuting operators
\begin{equation}
{\cal K}^{(J)}\equiv \nabla^a K^{(J)}_{ab}\nabla^b
\end{equation}
with the Klein--Gordon operator $\nabla^a g_{ab}\nabla^b$ for scalars. That is,
\be
\Big[{\cal K}^{(J)}, \nabla^a g_{ab}\nabla^b\Big]=0\,.
\ee
Thus, we have now $d$ commuting operators for the Klein--Gordon equation (including ${\cal L}_j=\xi^{(j)}_a\nabla^a$ for the Killing vectors $\xi^{(j)}=\partial_t,\partial_{\phi_j}$). Therefore we know the Hamilton--Jacobi and Klein--Gordon equations separate in these spacetimes~\cite{Frolov:2017kze}.

\section{Field Equations and Miscellanea}\label{AppC}
In this appendix we gather formulae regarding the quartic higher curvature gravities and their equations. We start with listing the 26 quartic curvature invariants.
\begin{align}
\mathcal{R}^{(4)}_{1} &= R_{a}{}^{e}{}_{c}{}^{f} R^{abcd} R_{e}{}^{j}{}_{b}{}^{h} R_{fjdh} 
\,, \quad 
\mathcal{R}^{(4)}_{2}=R_{a}{}^{e}{}_{c}{}^{f} R^{abcd} R_{bjdh} R_{e}{}^{j}{}_{f}{}^{h}
\,, \nn\\
\mathcal{R}^{(4)}_{3}&= R_{ab}{}^{ef} R^{abcd} R_{c}{}^{j}{}_{e}{}^{h} R_{djfh}
\,, \quad
\mathcal{R}^{(4)}_{4}=R_{ab}{}^{ef} R^{abcd} R_{ce}{}^{jh} R_{dfjh}
\,,\nn\\
\mathcal{R}^{(4)}_{5}&=R_{ab}{}^{ef} R^{abcd} R_{cdjh} R_{ef}{}^{jh}
\,, \quad
\mathcal{R}^{(4)}_{6}= R_{abc}{}^{e} R^{abcd} R_{fhjd} R^{fhj}{}_{e}
\,, 
\nonumber\\
\mathcal{R}^{(4)}_{7}&=(R_{abcd} R^{abcd})^2
\,,\quad
\mathcal{R}^{(4)}_{8}= R^{ab} R_{c}{}^{h}{}_{ea} R^{cdef} R_{dhfb}
\,, \quad
\mathcal{R}^{(4)}_{9}=R^{ab} R_{cd}{}^{h}{}_{a} R^{cdef} R_{efhb} 
\,,
\nonumber\\
\mathcal{R}^{(4)}_{10}&=R^{ab} R_{a}{}^{c}{}_{b}{}^{d} R_{efhc} R^{efh}{}_{d}
\,, \quad
\mathcal{R}^{(4)}_{11}= R R_{a}{}^{c}{}_{b}{}^{d} R_{c}{}^{e}{}_{d}{}^{f} R_{e}{}^{a}{}_{f}{}^{b}
\,, \quad 
\mathcal{R}^{(4)}_{12}=R R_{ab}{}^{cd} R_{cd}{}^{ef} R_{ef}{}^{ab},
\,
\nonumber\\
\mathcal{R}^{(4)}_{13}&= R^{ab} R^{cd} R_{ebfd} R^{e}{}_{a}{}^{f}{}_{c}
\,, \quad
\mathcal{R}^{(4)}_{14}=R^{ab} R^{cd} R_{ecfd} R^{e}{}_{a}{}^{f}{}_{b}
\,, \quad
\mathcal{R}^{(4)}_{15}= R^{ab} R^{cd} R_{efbd} R^{ef}{}_{ac}
\,,
\nonumber\\
\mathcal{R}^{(4)}_{16}&= R^{ab} R_{b}{}^{c} R_{defc} R^{def}{}_{a}
\,, \quad 
\mathcal{R}^{(4)}_{17}=R_{ef} R^{ef} R_{abcd} R^{abcd}
\,,\quad
\mathcal{R}^{(4)}_{18}=R^{de} R R_{abcd} R^{abc}{}_{e}
\,, 
\nonumber\\
\mathcal{R}^{(4)}_{19}&= R^2 R_{abcd} R^{abcd}
\, , \quad  
\mathcal{R}^{(4)}_{20}=R^{ab} R_{e}{}^{d} R^{ec} R_{acbd}
\,, \quad
\mathcal{R}^{(4)}_{21}=R^{ac} R^{bd} R R_{abcd}
\, ,  
\nonumber\\ 
\mathcal{R}^{(4)}_{22}&=R_{a}{}^{b} R_{b}{}^{c} R_{c}{}^{d} R_{d}{}^{a}
\,, \quad
\mathcal{R}^{(4)}_{23}=(R_{ab} R^{ab})^2 \,, \quad 
\mathcal{R}^{(4)}_{24}=R_{a}{}^{b} R_{b}{}^{c} R_{c}{}^{a} R
\,,
\nonumber\\
\mathcal{R}^{(4)}_{25}&=R_{ab} R^{ab} R^2
\,, \quad  \mathcal{R}^{(4)}_{26}=R^4 \, .
\end{align}
The corresponding field equations are written simply in terms of the following tensor:
\be 
P_{abcd} = g_{a[c} g_{d] b} + \sum_{i} \alpha_i P_{(i)}{}_{abcd}^{(2)} + \sum_{i} \beta_i P_{(i)}{}_{abcd}^{(3)} + \sum_{i} \gamma_i P_{(i)}{}_{abcd}^{(4)}\,,
\ee
where we will now list the particular cases. To simplify the notation in the more complicated cases, we introduce the following anti-symmetrization operator:
\be 
\tensor*{A}{_{abcd}^{lmop}} = \frac{1}{2} \delta_{[a}^{l} \delta_{b]}^{m} \delta_{[c}^{o} \delta_{d]}^{p}  + \frac{1}{2} \delta_{[a}^{o} \delta_{b]}^{p} \delta_{[c}^{l} \delta_{d]}^{m} \,, 
\ee
which satisfies
\be 
\tensor*{A}{_{abcd}^{lmop}} R_{lmop} = R_{abcd} \, .
\ee
Our conventions are standard, i.e. $
B_{[a|c|b]} = \frac{1}{2} B_{acb} - \frac{1}{2} B_{bca} \, .$
\begin{align}
P_{(1)}{}_{abcd}^{(2)} =& 2 R_{abcd} \, ,\quad 
P_{(2)}{}_{abcd}^{(2)} = \frac{1}{2} g_{bd} R_{ac} -  \frac{1}{2} g_{bc} R_{ad} -  \frac{1}{2} g_{ad} R_{bc} + \frac{1}{2} g_{ac} R_{bd} \, ,\nn
\\
P_{(3)}{}_{abcd}^{(2)} =& 2 g_{a[c} g_{d] b} R\,,
\quad
P_{(1)}{}_{abcd}^{(3)} =   3 R_{a}{}^{e}{}_{[c}{}^{f} R_{d]fbe} \, , \quad 
P_{(2)}{}_{abcd}^{(3)} = 3 R_{ab}{}^{ef} R_{cdef} \, ,\nn
\\
P_{(3)}{}_{abcd}^{(3)} =&  \frac{1}{16}\tensor*{A}{_{abcd}^{lmop}} \big[2 R_{p}{}^{e} R_{lmoe} +  g_{mp} R_{l}{}^{efh} R_{oefh}  \big] \, ,
\nn\\
P_{(4)}{}_{abcd}^{(3)} =&   g_{a[c} g_{d] b} R_{efgh}R^{efgh} +  R P_{(1)}{}_{abcd}^{(2)}  \, ,\nn   
\\
P_{(5)}{}_{abcd}^{(3)} =& \frac{1}{16}\tensor*{A}{_{abcd}^{lmop}}  \big[R_{lo} R_{mp} + 2 g_{lo} R^{ef} R_{mepf} \big] \,,\nn\\
P_{(6)}{}_{abcd}^{(3)} =& \frac{3}{4} g_{bd} R_{a}{}^{e} R_{ce} -  \frac{3}{4} g_{ad} R_{b}{}^{e} R_{ce} -  \frac{3}{4} g_{bc} R_{a}{}^{e} R_{de} + \frac{3}{4} g_{ac} R_{b}{}^{e} R_{de} \,,\nn
\\
P_{(7)}{}_{abcd}^{(3)} =& g_{a[c} g_{d] b} R_{ef}R^{ef} + R P_{(2)}{}_{abcd}^{(2)} \, ,
\quad 
P_{(8)}{}_{abcd}^{(3)} = 3 g_{a[c} g_{d] b} R^2 \, .
\end{align}
\begin{align}
P_{(1)}{}_{abcd}^{(4)} =& 4 R_{[a}{}^h{}_{|e|}{}^i R_{b]}{}^e{}_{[d}{}^f R_{c]hfi} \, ,\quad
P_{(2)}{}_{abcd}^{(4)} =  4 R_{a}{}^e{}_{[c}{}^f R_{|b|}{}^h{}_{d]}{}^i R_{ehfi} \, , \nn
\\
P_{(3)}{}_{abcd}^{(4)} =& \frac{1}{8}\tensor*{A}{_{abcd}^{lmop}}\tensor{R}{_l^e_o^f}\tensor{R}{_{me}^{hi}} \tensor{R}{_{pfhi}} + \tensor{R}{_a^h_e^i}\tensor{R}{_{cd}^{ef}}\tensor{R}{_{bhfi}} + \tensor{R}{_c^h_e^i}\tensor{R}{_{ab}^{ef}} \tensor{R}{_{dhfi}} \, ,\nn
\\
P_{(4)}{}_{abcd}^{(4)} =& 2 R_{ae}{}^{hi} R_{bfhi} R_{cd}{}^{ef} + 2 R_{ab}{}^{ef} R_{ce}{}^{hi} R_{dfhi} \, ,\quad
P_{(5)}{}_{abcd}^{(4)} = 4 R_{ab}{}^{ef} R_{cd}{}^{hi} R_{efhi} \, ,\nn\\
P_{(6)}{}_{abcd}^{(4)} =& \frac{1}{4}\tensor*{A}{_{abcd}^{lmop}} \tensor{R}{_l^e_{op}} \tensor{R}{_m^{fhi}} \tensor{R}{_{efhi}} \, , \quad 
P_{(7)}{}_{abcd}^{(4)} = 4 R_{abcd} R_{efhi} R^{efhi} \, ,\nn 
\\
P_{(8)}{}_{abcd}^{(4)} =&  \frac{1}{16}\tensor*{A}{_{abcd}^{lmop}} \big[ \tensor{R}{^{ef}} \tensor{R}{_{leo}^h} \tensor{R}{_{m f p h}} + 2 \tensor{R}{_p^e}\tensor{R}{_l^f_o^h}\tensor{R}{_{mfeh}} +  g_{m p} \tensor{R}{_l^{efh}} \tensor{R}{_o^i_f^j}\tensor{R}{_{eihj}} \big] \, ,\nn
\\
P_{(9)}{}_{abcd}^{(4)} =&  \frac{1}{16}\tensor*{A}{_{abcd}^{lmop}} \big[2\tensor{R}{_m^e}\tensor{R}{_{lefh}} \tensor{R}{_{op}^{fh}} - g_{l p} \tensor{R}{_m^{efh}} \tensor{R}{_{oe}^{ij}} \tensor{R}{_{fhij}} \big] \, ,\nn
\\
P_{(10)}{}_{abcd}^{(4)} =& \frac{1}{16}\tensor*{A}{_{abcd}^{lmop}} \big[  g_{mp} R_{e}{}^{hij} R_{fhij} R_{l}{}^{e}{}_{o}{}^{f} + 2 R^{ef} R_{lefh} R_{m}{}^{h}{}_{op} + R_{mp} R_{l}{}^{efh} R_{oefh}\big] \,, 
\nn\\
P_{(11)}{}_{abcd}^{(4)} =&  g_{a[c} g_{d] b} \mathcal{R}^{(3)}_{1} + R P_{(1)}{}_{abcd}^{(3)} \, ,\quad 
P_{(12)}{}_{abcd}^{(4)} =  g_{a[c} g_{d] b} \mathcal{R}^{(3)}_{2} + R P_{(2)}{}_{abcd}^{(3)}\,,\nn
\\
P_{(13)}{}_{abcd}^{(4)} =&  \frac{1}{8}\tensor*{A}{_{abcd}^{lmop}} \big[R_{m}{}^{e} R_{p}{}^{f} R_{leof} + g_{mp} R^{ef} R_{l}{}^{h}{}_{e}{}^{i} R_{ohfi} \big] \, ,\nn
\\
P_{(14)}{}_{abcd}^{(4)} =&  \frac{1}{8}\tensor*{A}{_{abcd}^{lmop}} \big[ R^{ef} R_{mp} R_{leof} + g_{mp} R^{ef} R_{ehfi} R_{l}{}^{h}{}_{o}{}^{i} \big] \, ,\nn
\\
P_{(15)}{}_{abcd}^{(4)} =& \frac{1}{8}\tensor*{A}{_{abcd}^{lmop}} \big[g_{mp} R^{ef} R_{le}{}^{hi} R_{ofhi} + R_{l}{}^{e} R_{m}{}^{f} R_{opef} \big] \, ,\nn
\\
P_{(16)}{}_{abcd}^{(4)} =& \frac{1}{8}\tensor*{A}{_{abcd}^{lmop}}  \big[g_{mp} R_{o}{}^{e} R_{efhi} R_{l}{}^{fhi} + R_{e}{}^{f} R_{p}{}^{e} R_{lmof} \big] \, ,\nn
\\
P_{(17)}{}_{abcd}^{(4)} =& P_{(1)}{}_{abcd}^{(2)} \mathcal{R}^{(2)}_{2} +\mathcal{R}^{(2)}_{1}  P_{(2)}{}_{abcd}^{(2)}  \, ,\quad
P_{(18)}{}_{abcd}^{(4)} =  g_{a[c} g_{d] b}  \mathcal{R}^{(3)}_{3} + R P_{(3)}{}_{abcd}^{(3)}  \, ,\nn
\\
P_{(19)}{}_{abcd}^{(4)} =& P_{(1)}{}_{abcd}^{(2)} \mathcal{R}_{3}^{(2)} +   \mathcal{R}_{1}^{(2)} P_{(3)}{}_{abcd}^{(2)} \, ,\nn
\\ 
P_{(20)}{}_{abcd}^{(4)} =& \frac{1}{16}\tensor*{A}{_{abcd}^{lmop}} \big[ R_{l}{}^{e} R_{mp} R_{oe} + 2 g_{mp} R^{fh} R_{o}{}^{e} R_{lfeh} +  g_{mp} R_{e}{}^{h} R^{ef} R_{lfoh} \big] \, ,\nn
\\
P_{(21)}{}_{abcd}^{(4)} =&  g_{a[c} g_{d] b} \mathcal{R}^{(3)}_{5} + R P_{(5)}{}_{abcd}^{(3)} \, ,\quad 
P_{(22)}{}_{abcd}^{(4)} = \frac{1}{4}\tensor*{A}{_{abcd}^{lmop}} \ g_{mp} R_{ef} R_{l}{}^{e} R_{o}{}^{f} \, ,\nn
\\
P_{(23)}{}_{abcd}^{(4)} =& 2 \mathcal{R}_{2}^{(2)} P_{(2)}{}_{abcd}^{(2)} \, ,\quad 
P_{(24)}{}_{abcd}^{(4)} =  g_{a[c} g_{d] b}  \mathcal{R}^{(3)}_{6} + R P_{(6)}{}_{abcd}^{(3)}  \, ,\nn
\\
P_{(25)}{}_{abcd}^{(4)} =& 2 g_{a[c} g_{d] b} R (\mathcal{R}_{2}^{(2)})^2 + 2 R^2 P_{(2)}{}_{abcd}^{(2)} \, ,\quad
P_{(26)}{}_{abcd}^{(4)} = 4 g_{a[c} g_{d] b} R^3 \, .
\end{align}

\bibliography{references}
\bibliographystyle{JHEP}

\end{document}